\documentclass[12pt,epsf]{article}

\usepackage{scicite}

\input psfig.sty

\def\gsim{\mathrel{\rlap{\lower 4pt \hbox{\hskip 1pt $\sim$}}\raise 1pt
\hbox {$>$}}}
\def\lsim{\mathrel{\rlap{\lower 4pt \hbox{\hskip 1pt $\sim$}}\raise 1pt
\hbox {$<$}}}

\newenvironment{sciabstract}{%
\begin{quote} \bf}
{\end{quote}}

\newcounter{lastnote}

\newcommand{\HII}{H{\sc ii} }

\title{Structure Formation in the Early Universe} 
\author{
Naoki Yoshida,$^{1}$\\
\normalsize{$^{1}$Institute for the Physics and Mathematics of the Universe,}\\
\normalsize{University of Tokyo}\\
\normalsize{5-1-5 Kashiwanoha, Kashiwa, Chiba 277-8568, Japan}\\
\\
\normalsize{$^\ast$To whom correspondence should be addressed; E-mail:
naoki.yoshida@ipmu.jp} \\
\normalsize{{\bf To appear in Advanced Science Letters, 2009}}
}

\date{}

\begin{document} 


\baselineskip24pt


\maketitle


\begin{sciabstract}
The standard theory of cosmic structure formation posits that
the present-day rich structure of the Universe developed through
gravitational amplification of tiny matter density fluctuations generated
in its very early history. 
Recent observations of the cosmic microwave
background, large-scale structure, and distant supernovae
determined the energy content of the Universe and the
basic statistics of the initial density field with great accuracy. 
It has become possible to make accurate
predictions for the formation and nonlinear growth of structure 
from early to the present epochs.
We review recent progress in the theory of structure formation in 
the universe. We focus on the formation of the first cosmological objects.
Results from state-of-the-art numerical simulations are presented.
Finally, we discuss prospects for future observations of the first 
generation of stars and galaxies.
\end{sciabstract}

-- cosmology, star formation, dark matter --

\section{Introduction: The Dark Ages}
Rich structures in the universe we see today, 
such as galaxies and galaxy clusters, have developed over a very long time.
Astronomical observations utilizing 
large ground-based telescopes discovered distant  
galaxies and quasars\cite{SDSS, Iye} 
that were in place when the Universe was less than one billion years old.  
We can probe directly, although not completely, 
the evolution of the cosmic structure all the way from the present-day to 
such an early epoch. 
We can also observe the state of the Universe at an even earlier epoch, about 380,000
years after the Big Bang, as the cosmic microwave
background (CMB). The anisotropies of CMB provide information on the
{\it initial conditions} for the formation of all the structures.
In between these two epochs lies the remaining frontier of astronomy, 
when the universe was about a few to several
million years old. The epoch is called the cosmic Dark Ages\cite{Miralda03}. 

Shortly after the cosmological recombination epoch when
hydrogen atoms were formed and the CMB photons were last-scattered, 
the CMB shifted to infrared, and then the universe would have appeared completely dark
to human eyes. A long time had to pass until the first stars were born,
which illuminate the universe once again and terminate the Dark Ages.
The first stars are thought to be the first sources of light,
and also the first sources of heavy elements that enable the 
formation of ordinary stellar populations, planets, and ultimately,
the emergence of life.

Over the past years, there have been a number of theoretical studies on the yet unrevealed
era in the cosmic history. 
Perhaps the current particular fascination 
with such studies is owing to recent rapid progress in observational astronomy.
Existing telescopes are already probing early epochs
close to the end of the Dark Ages, and planned observational
programs are aimed at detecting directly light from objects farther away.

In this article, we review recent progress in the theory of structure formation in 
the early universe. 
We focus on the formation of 
the first generation stars.
Theoretical studies hold promise for revealing the detailed process
of primordial star formation for two main reasons: (1) the initial
conditions, as determined cosmologically, are well-established, so
that statistically equivalent realizations of a standard model
universe can be accurately generated, and (2) all the important basic
physics such as gravitation, hydrodynamics, and atomic and molecular
processes in a hydrogen-helium gas are understood.  
In principle, therefore, it is possible to make solid 
predictions for the formation of early structure and of the first stars
in an expanding universe. We describe some key physical processes.
Computer simulations are often used to tackle the nonlinear
problems of structure formation. We show the results from large cosmological 
$N$-body hydrodynamic simulations.

\section{Hierarchical structure formation and the first cosmological objects}
We first describe generic hierarchical nature of structure formation
in the standard cosmological model which is based on weakly-interacting
Cold Dark Matter (CDM). 
The primordial density fluctuations predicted by popular inflationary universe models
have very simple characteristics\cite{Riotto}. 
They are described by a Gaussian random
field, and have a nearly scale-invariant power spectrum $P(k)\propto k^n$ for
wavenumber $k$ with
$n\sim 1$. 
The perturbations come into the horizon 
in the radiation-dominated and then in the matter-dominated epochs.
The difference in the growth rate in these epochs 
results in a modified power spectrum\cite{Dodelson}. 
Effectively the slope of the power-spectrum
changes slowly as a function of length scale, but the final shape is still
simple and monotonic in CDM models. 
The CDM density fluctuations have progressively larger amplitudes
on smaller length scales. Hence structure formation is expected
to proceed in a ``bottom-up'' manner, with smaller objects forming earlier.

\begin{figure}
\begin{center}
\psfig{file=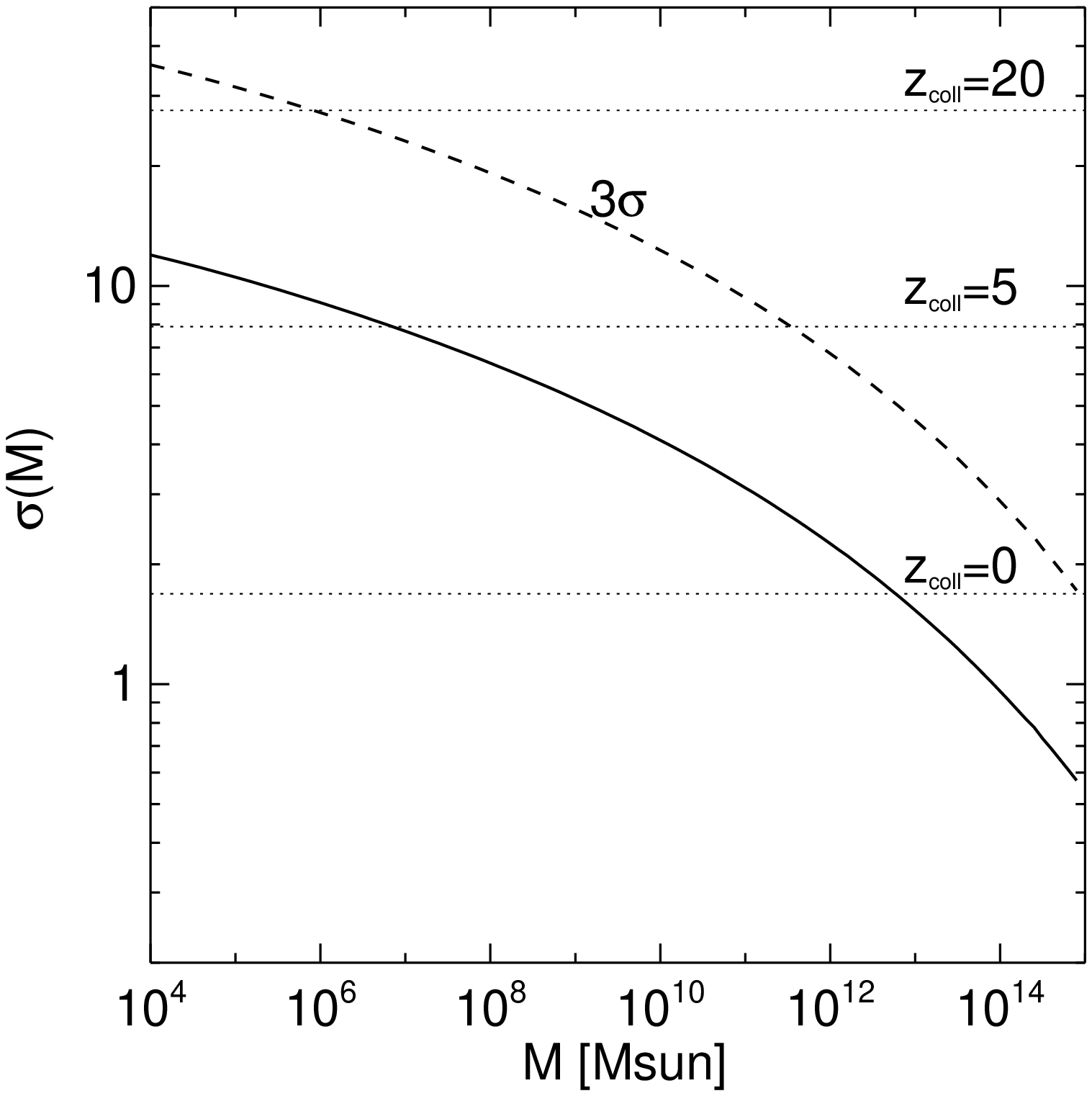,width=1\textwidth}
  \caption{Mass variance and collapse thresholds 
for a flat CDM model with cosmological constant.
The assumed cosmological parameters are, matter density $\Omega_{\rm m}$ = 0.3,
baryon density $\Omega_{\rm b}$ = 0.04, amplitude of fluctuations
$\sigma_8 = 0.9$, and the Hubble constant $H_0 = 70$ km s$^{-1}$ Mpc$^{-1}$.
}
\label{fig:sigma}
\end{center}
\end{figure}

It is useful to work with a properly defined mass variance
to obtain the essence of hierarchical structure formation.
The mass variance is defined as the root-mean square of mass density
fluctuations within a sphere that contains mass $M$. It is given
by a weighted integral of the power spectrum as 
\begin{equation}
\sigma^2 (M)=\frac{1}{2\pi^2}\int P(k) W^2(k R) k^2 {\rm d}k,
\label{eq:variance}
\end{equation}
where the top-hat window function is given by 
$W(x)=3(\sin (x)/x^3 - \cos (x)/x^2)$. 
Let us define the threshold over-density for gravitational collapse
at redshift $z$ as
\begin{equation}
\delta_{\rm crit} (z)=1.686/D(z),
\end{equation}
where $D(z)$ is the linear growth factor of perturbations to $z$.
Fig.~\ref{fig:sigma} show the variance and the collapse threshold at
$z=0,5,20$. At $z=20$, the mass of a halo which corresponds to a 
3-$\sigma$ fluctuation is just about $10^6 M_{\odot}$.
As shown later in section \ref{sec:early}, this is
the characteristic mass of the first objects in which the primordial 
gas can cool and condense by molecular hydrogen cooling. 

The mass variance is sensitive to the shape of the initial power spectrum. 
For instance, in warm dark matter models in which 
the power spectrum has an exponential cut-off at the particle
free-streaming scale, the corresponding mass variance 
at small mass scales is significantly reduced\cite{HaimanBarkana, YoshidaWarm}.
In such models, early structure formation is effectively delayed, 
and hence small nonlinear objects form later than in the CDM model.
Thus the formation epoch of the first objects and hence the beginning of 
cosmic reionization have a direct link to 
the nature of dark matter and the shape of the primordial density fluctuations
\cite{Kosowsky, Somerville, YoshidaRSI}.
We discuss this issue further in Section \ref{sec:DM}.

\section{Formation of the first cosmological objects\label{sec:early}}
The basics of the formation of nonlinear dark matter halos are easily understood; 
because of its hierarchical nature, dark matter
halos form in essentially the same way regardless of mass and the formation
epoch. Halos will form at all mass scales by gravitational instability from 
scale-free density fluctuations.
The first 'dark' objects are well defined, and
are indeed halos of a very small mass which is set by
dark matter particles' initial free-streaming motion\cite{Diemand}. 

The formation of the first baryonic objects 
involves a number of physical processes, and so is much more complicated.
Baryons can collapse in dark matter halos only if the radiative cooling time
is shorter than the age of the universe\cite{Ostriker}.
The study of the evolution of primordial gas in the early universe 
and the origin of the first
baryonic objects has a long history\cite{Matsuda,Kashlinsky,
Couchman}. The emergence of the standard cosmological model has enabled 
us to ask more specific questions, such as {\it when did the first objects
form}, and {\it what is the characteristic mass} ?
  
\begin{figure}
\psfig{file=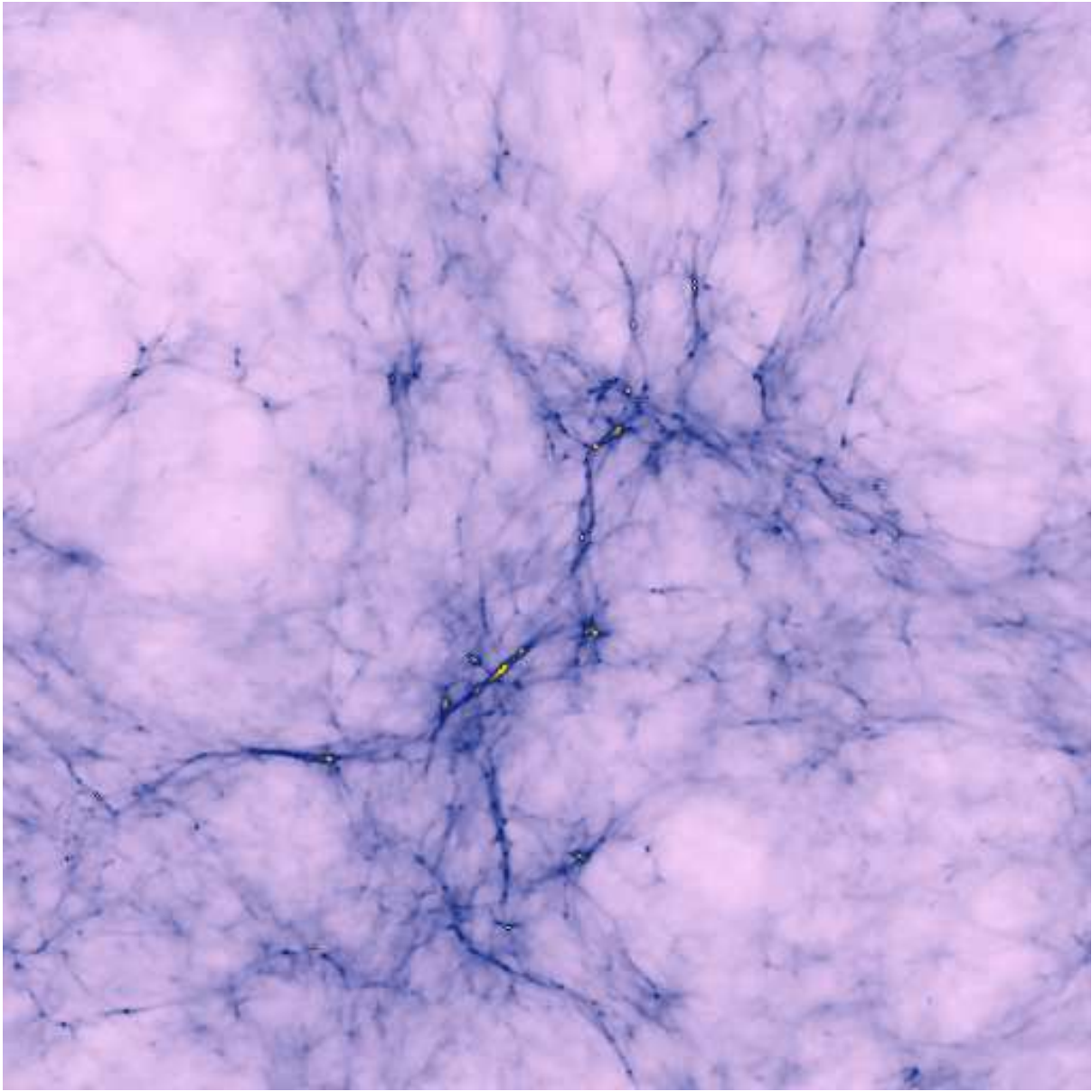,width=1\textwidth}
  \caption{The projected gas distribution at $z=17$ in a cubic volume of 600$h^{-1}$kpc 
on a side. The cooled dense gas clouds appear as bright spots at the intersections 
of the filamentary structures. From Ref. \cite{YoshidaFirst}.}
\label{fig:first}
\end{figure}

Recent numerical simulations of early structure formation
show that this process likely began as early
as when the age of the universe is less than a million years\cite{Miralda03, Gao07}.
In these simulations, dense, cold clouds of self-gravitating
molecular gas develop in the inner regions of small dark halos and contract
into proto-stellar objects with masses in the range $\sim 100 - 1000
M_{\odot}$. Fig.~\ref{fig:first} shows the projected gas distribution 
in a cosmological simulation that includes hydrodynamics and 
primordial gas chemistry\cite{YoshidaFirst}.
Star-forming gas clouds are found at
the knots of filaments, resembling the large-scale
structure of the universe, although actually much smaller in mass and size. 
This manifest the hierarchical nature of structure in the CDM universe.

 Unlike the formation of dark matter halos which is solely governed by gravity,
star formation involves at least several major processes as follows.
For star formation to begin in the early universe,
a sufficient amount of cold dense gas must accumulate in a dark halo.
The primordial gas cannot efficiently cool radiatively because
atoms have excitation energies that are too high, and molecules, which have accessible
rotational energies, are very rare. Trace amounts of molecular hydrogen (${\rm H}_2$) can form via
a sequence of reactions,
\begin{equation}
{\rm H} + e^{-} \rightarrow {\rm H}^{-} + \gamma,
\label{eq:Hminus}
\end{equation}
\begin{equation}
{\rm H}^{-} + {\rm H} \rightarrow {\rm H}_2 + e^{-}. 
\label{eq:form2}
\end{equation}
H$_2$ molecules, once formed, can change their quantum rotational
and vibrational levels and emit photons. This allows the
gas to cool and eventually condense to form gas clouds.

The critical temperature for these processes to operate is found to be about 2000 K.
We here follow Ref. \cite{Tegmark97}
to derive this characteristic temperature.
Consider a gas with particle number density $n$ and temperature $T$.
The ionization fraction $x = n[{\rm H}^{+}]/n$ 
and the molecular fraction $f = n[{\rm H}_{2}]/n$ evolve as
\begin{equation}
\dot{x} = -k_{\rm rec} \; n \; x^2,
\label{eq:recom}
\end{equation}
\begin{equation}
\dot{f} = k_{\rm form} \; n \; (1-x-2f) \; x,
\label{eq:H2form}
\end{equation}
where $k_{\rm rec}$ is hydrogen recombination rate
and $k_{\rm form}$ is the net formation rate of molecular hydrogen.
Equation (\ref{eq:recom}) simply describes the rate of hydrogen recombination,
by which the ionization fraction decreases, and equation (\ref{eq:H2form})
gives the rate of H$_2$ formation determined mainly by 
the above reaction (\ref{eq:form2}).
Assuming the cloud density and temperature remain
roughly constant in a virialized halo, and taking $1-x-2f \sim 1$
for a neutral cosmic primordial gas, we obtain solutions
\begin{equation}
x (t) = \frac{x_0}{1 + x_0\, n \,k_{\rm rec}\, t}\;,
\end{equation}
\begin{equation}
f (t) = f_0 + \frac{k_{\rm form}}{k_{\rm rec}}
\ln (1 + x_0 \, n \, k_{\rm rec} \, t).
\end{equation}
Note the logarithmic dependence of $f$ on time.
Substituting the temperature dependence of the reaction rates ${k_{\rm rec}}, {k_{\rm form}}$, 
we obtain, after some straightforward algebra, 
a simple scaling of
an asymptotic molecular fraction
\begin{equation}
f_{\rm c} \propto T^{1.52} .
\label{eq:asym}
\end{equation}

\begin{figure}
\psfig{file=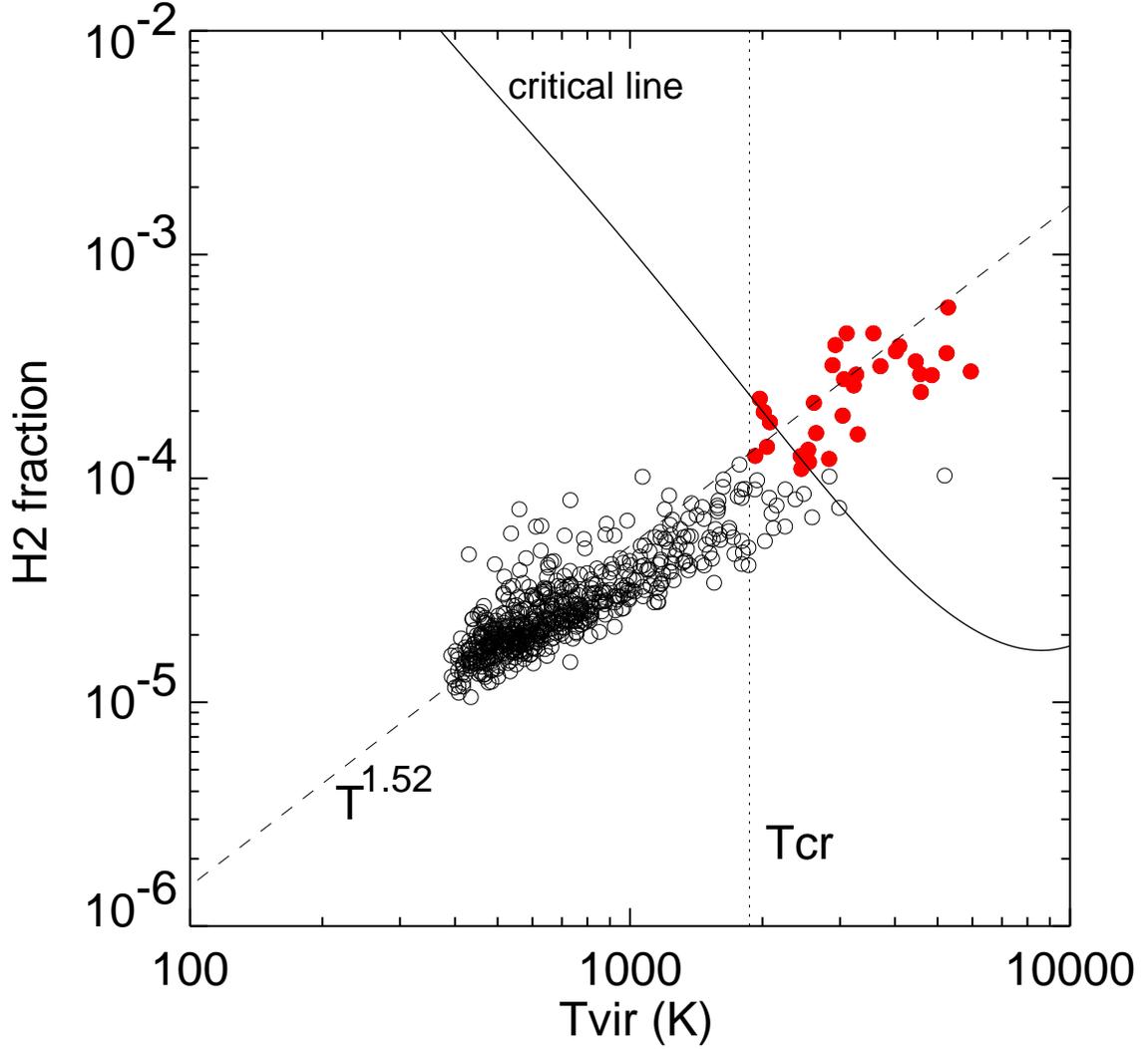,width=1\textwidth}
  \caption{The mass weighted mean H$_{2}$ fraction versus virial temperature 
for the halos that host gas clouds (filled circles) and for those that do not 
(open circles) at $z=17 (t_{\rm age} = 300$ Myrs). 
The solid curve is the H$_{2}$ fraction needed to cool the gas
at a given temperature and the dashed line is the asymptotic H$_{2}$
fraction (see equation [\ref{eq:asym}]). From Ref. \cite{YoshidaFirst}.}
  \label{fig:first2}
\end{figure}

Remarkably, this simple scaling is shown to provide a rather accurate estimate.
Fig.~\ref{fig:first2} shows the molecular fraction $f_{\rm H_{2}}$ 
against the virial temperature for halos located in a large
cosmological simulation.
The solid line is an analytical estimate of the H$_{2}$ fraction needed to
cool the gas. It is computed from the cooling function of H$_2$ molecules
\cite{GP98}.
In Fig.~\ref{fig:first2}, halos appear to be clearly separated into two
populations; those in which the gas has cooled (solid circles), and the
others (open circles).  
The analytic estimate yields a critical temperature of $\sim 2000$ K,
which indeed agrees very well
with the distribution of gas in the $f_{\rm H_{2}}$ - $T$ plane.
There is an important dynamical effect, however.
The gas in halos
that accrete mass rapidly (primarily by mergers) is unable to cool
efficiently owing to gravitational and gas dynamical heating. 
The effect explains the spread of halos into two populations
at $T \sim 2000-5000$ K. Therefore, ``minimum collapse mass''
models are a poor characterization of primordial gas cooling and gas
cloud formation in the hierarchical CDM model.
The formation process is significantly affected 
by the dynamics of gravitational collapse.
It is important to take into account the details of halo formation 
history \cite{YoshidaFirst, Reed}.

\section{The role of  dark matter and dark energy}
\label{sec:DM}

The basic formation process of the first objects
is described largely by the physics of a primordial gas.
Its thermal and chemical evolution specifies a few important
mass scales, such as the 
Jeans mass at the onset of collapse (see Section 5). 
However, when and how primordial gas clouds are formed are
critically affected by the particle properties of dark matter,
by the shape and the amplitude of the initial density
perturbations, and by the overall expansion history of the universe.
We here introduce two illustrative
examples; a model in which dark matter is assumed to be
``warm'', and another cosmological model 
in which dark energy obeys a time-dependent equation of state. 

\begin{figure}
\psfig{file=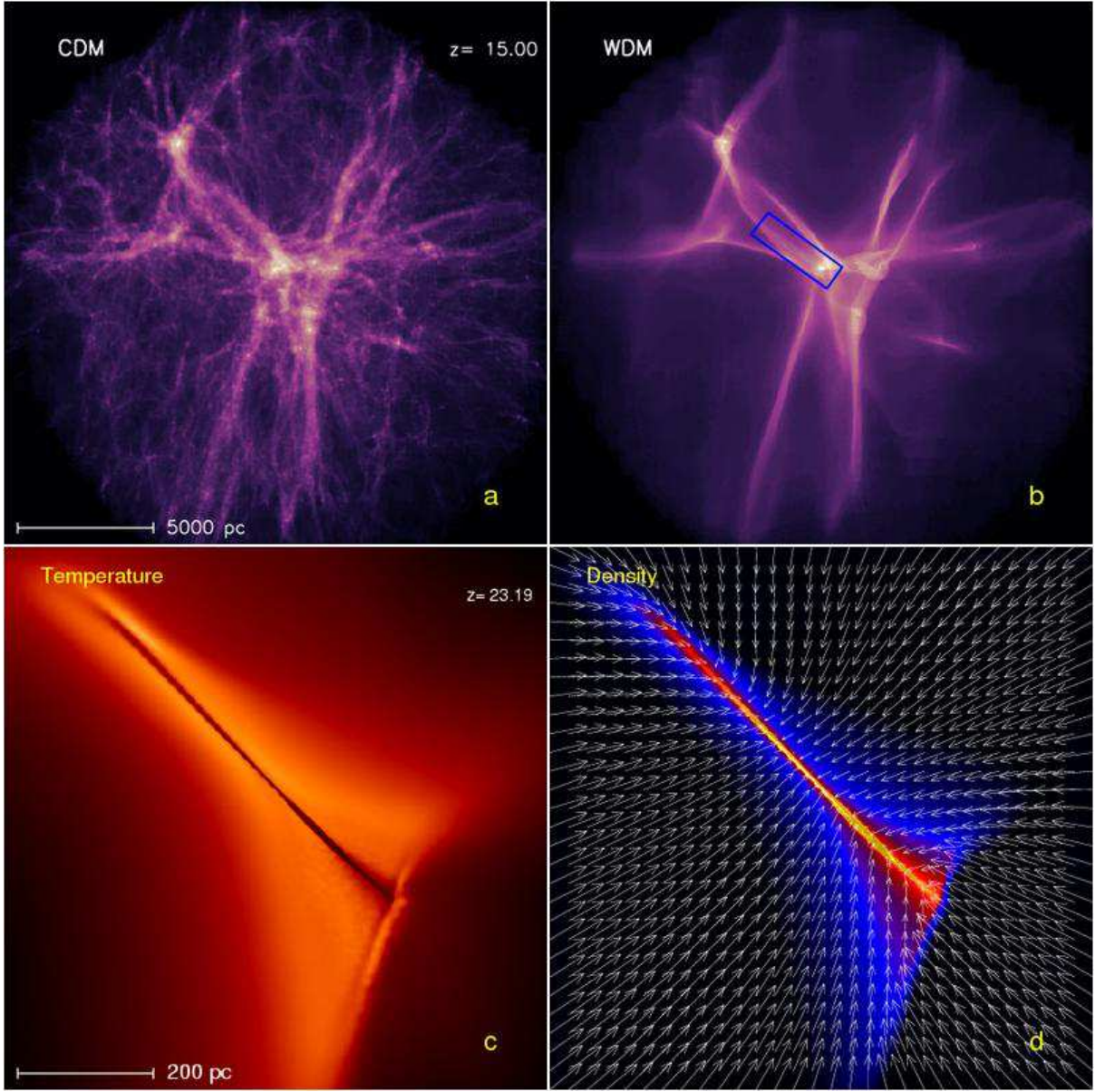,width=1\textwidth}
  \caption{The projected gas distribution at $z=15$ for
the standard CDM model (left) and for a WDM model (right).
We see much smoother matter distribution in the WDM model,
in which gas clouds are formed in the prominent filametary structure
(bottom panels). From Gao \& Theuns 2007, Science.}
  \label{fig:wdm}
\end{figure}

If dark matter is warm, the matter power spectrum has an exponential cut-off at the particle
free-streaming scale, and then the corresponding mass variance 
at small mass scales is significantly reduced\cite{HaimanBarkana, YoshidaWarm}.
The effect is clearly seen in Fig. \ref{fig:wdm}.
The gas distribution is much smoother in a model with 
warm dark matter. For the particular model with dark matter particle 
mass of 3 keV, dense gas clouds are formed in filamentary
shapes, rather than in blobs embedded in dark matter halos\cite{YoshidaWarm, wdmGao}.
While further evolution of the filamentary gas clouds
is uncertain, it is expected that stars are lined up along filaments.
Vigorous fragmentation of the filaments, if it occurs, can lead to
the formation of multiple low-mass stars.

Dark matter particles might affect primordial star formation in a very different way.
A popular candidate for dark matter is super-symmetric particles\cite{Jungman},
neutralinos for instance. Neutralinos are predicted to have a large cross-section
for pair-annihilation. 
Annihilation products are absorbed in a very dense gas clouds,
which can counteract molecular cooling\cite{Spolyar}.
Because primordial gas clouds are formed at the center
of dark matter halos, where dark matter density is very large,
the annihilation rate and resulting energy input can be significant.
While the net effect of dark matter annihilation
remains highly uncertain, it'd be interesting and even necessary
to include the effect if neutralinos are detected in laboratories.

\begin{figure}
\psfig{file=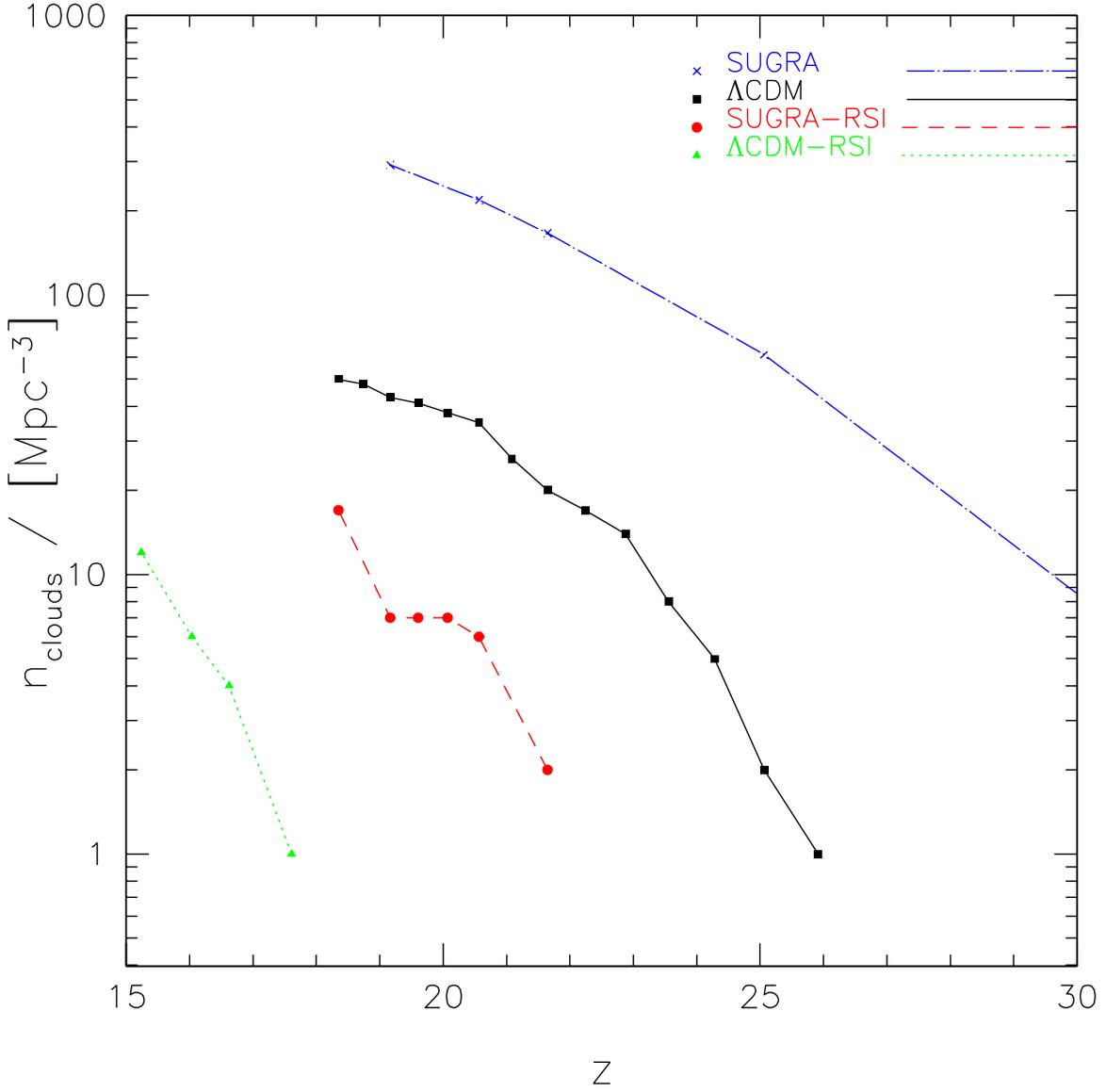,width=1\textwidth}
\caption{The number of primordial gas clouds
at high redshifts for a variety of models;
SUGRA (an evolving dark energy), $\Lambda$CDM,
SUGRA + a running spectral inflation model,
and $\Lambda$CDM + a running spectral inflation model.
From Ref. \cite{Maio}.}
  \label{fig:maio}
\end{figure}

The nature of dark energy also affects the formation epoch
of the first objects\cite{Maio}. The growth rate of density perturbations
is a function of cosmic expansion parameter, which is determined
by the energy content of the universe. 
In general, the energy density of dark energy can be written as 
\begin{equation}
\rho_{\rm DE} \propto \exp \left[ \int^{a} -3 \; \frac{da'}{a'} (1 + w (a')) \right],
\label{eq:darke}
\end{equation}
where $a$ is cosmic expansion parameter, and 
$w(a)$ defines the effective equation of state of dark energy
via $P = w \rho$.
For the simplest model
of dark energy, i.e., Einstein's cosmological constant with $w=-1$,
cosmic expansion is accelerated only at late epochs ($z<1$),
which is unimportant for early structure formation.
However, some dark energy models predict time-dependent
equation of state, which effectively shifts 
the formation epoch to early or later epochs.
Fig. \ref{fig:maio} shows the number of primordial gas clouds
as a function of redshift for the standard $\Lambda$CDM model
and for an evolving dark energy model.

Unfortunately, it is extremely difficult to measure 
the abundance of star-forming gas clouds as a function 
of time from currently available observations.
It is possible to infer how early cosmic reionization
began from the large-scale anisotropies of CMB polarization
\cite{Zaldarriaga,Komatsu08},
but the CMB polarization measurement does not put tight 
constraints on the reionization
history.
We will need to await for a long time until
future radio observations map out the distribution of the intergalactic medium
in the early universe
by detecting redshifted 21cm emission
from neutral hydrogen\cite{Shapiro06}. 

\section{Formation of the first stars}
Statistical properties of
primordial star-forming clouds and the overall effect of cosmological 
bias have been studied in detail \cite{YoshidaFirst, Gao07, Oshea07}.
We now describe more details of the formation process of stars -- the first stars.

The dynamics of primordial gas cloud collapse has been studied extensively 
over the past few decades\cite{Matsuda, PallaSalpeter83}.  
One-dimensional hydrodynamic simulations of spherical gas collapse were 
also performed with increasing levels of physics implementation\cite{ON98, Ripa02}.  
These studies showed that, while the overall evolution
can be understood using a self-similar collapse model\cite{Larson}, there are clear differences in the
thermal evolution of a primordial gas cloud from that of present-day, metal- and
dust-enriched gas clouds.
Three-dimensional cosmological simulations were performed
by several groups so far\cite{ABN, BCL}. 
These calculations achieve a large dynamic range and implement
primordial gas chemistry, 
and hence were able to follow the evolution of a primordial 
gas cloud in detail. They showed clearly how early gas clouds are 
formed in a cosmological context. However, the calculations are
stopped at intermediate
phases where the gas cloud is still gravitationally contracting.

Recently, an {\it ab initio} simulation of the formation of a primordial protostar
has been finally performed\cite{YOH08}.
The simulation has an extraordinary spatial resolution, 
so that the highest gas densities reach
``stellar'' density, and thus it 
offers a detailed picture of how the first cosmological objects,
protostars, form from primeval density fluctuations left over from the Big Bang.  
Unlike most simulations of star formation, the simulation does not assume 
any {\it a priori} equation of state for the gas. The thermal and chemical
evolution is fully determined by molecular and atomic
processes, including molecular hydrogen formation at both
low and high densities, and transfer of molecular lines and continuum radiation.
All of these processes are treated in a direct, self-consistent manner.

\begin{figure}
\begin{center}
\psfig{file=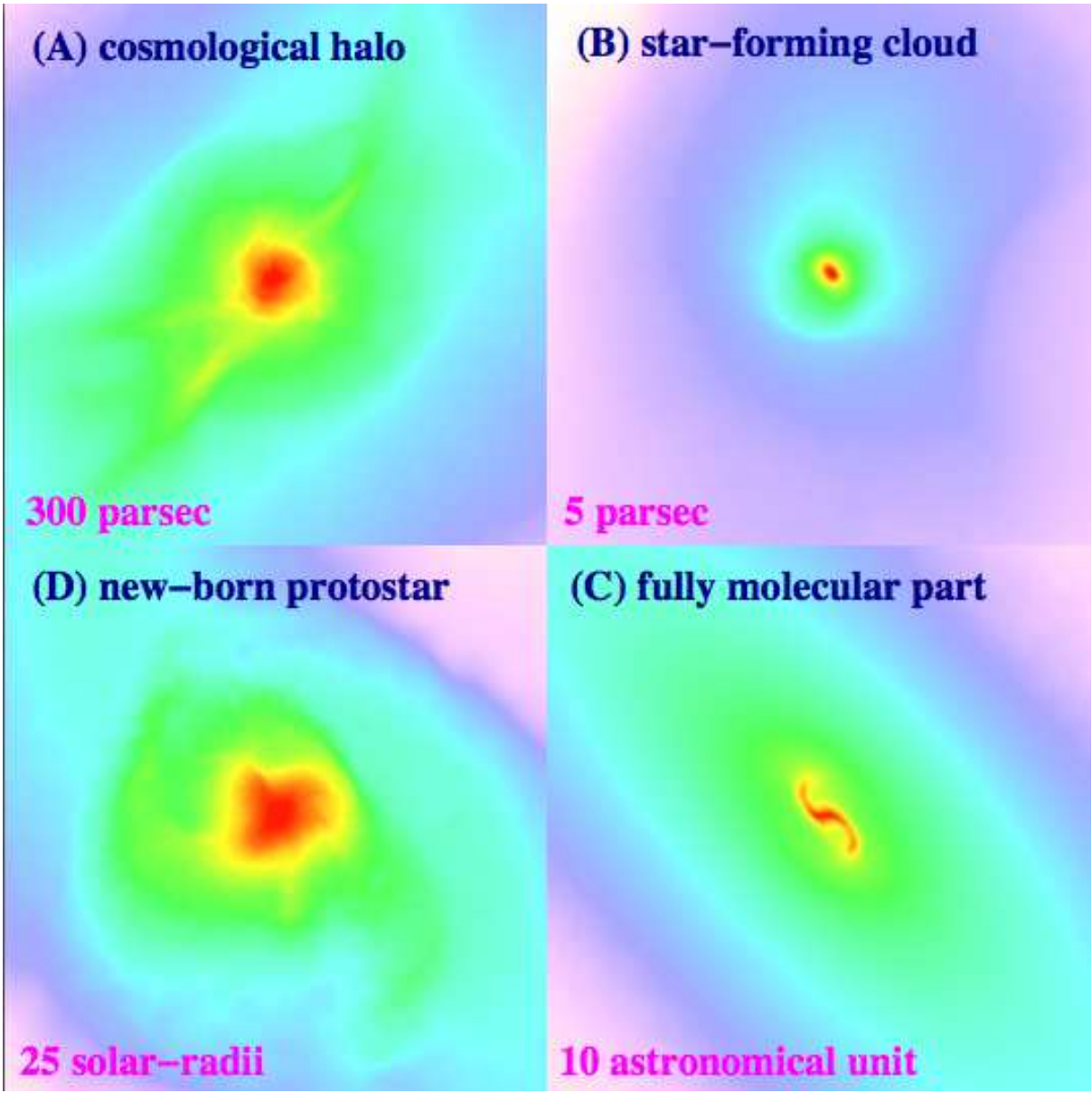,width=1\textwidth}
\caption{ Projected gas distribution around
the protostar. Shown regions are, from top-left, clockwise, 
the large-scale gas distribution around the cosmological halo
(300 pc on a side), a self-gravitating, star-forming cloud (5 pc on a side), 
the central part of the fully molecular core (10 astronomical units on a side),
and the final protostar (25 solar-radii on a side).
We use the density-weighted temperature to color the bottom-left panel,
to show clearly the complex structure of the protostar.
From Ref. \cite{YOH08}.}
\label{fig:proto}
\end{center}
\end{figure}

We describe in detail the simulation of Ref. \cite{YOH08}, 
which followed the gravitational collapse of dark matter and the hydrodynamics 
of primordial gas. 
Small dark halos of about a half million solarmasses are assembled when the age
of the universe is a few million years old. 
Through the action of radiative cooling, star-forming gas clouds
collect in their host dark halo. 
Fig.~\ref{fig:proto} shows the projected gas density in and around the prestellar
gas cloud. Note that the figures were made from a single simulation which
covers a very large dynamic range of $\sim 10^{13}$ in length scale.
We see substantial variations in density and temperature 
even in the innermost 10 solar-radii region around the newly formed
protostar. 

Through a number of atomic and molecular processes, 
the cloud contract roughly isothermally; the density increases
over 20 decades, but the temperature increase only by a factor
of 10. When the central density reaches $n\sim 10^{18}\; {\rm cm}^{-3}$, 
the gas becomes completely optically thick to continuum radiation, and 
radiative cooling does not operate efficiently any more.
Because the cloud core had initially a small angular momentum,
the central part flattens to form a disk-like structure at this point. 
Further collapse and the associated dynamical heating triggers 
full-scale dissociation of hydrogen molecules in the central part.  
Thereafter,
the gas cannot lose its thermal energy neither radiatively nor 
by dissociating molecules. 
The gas then contracts adiabatically, 
and its temperature quickly increases above several thousand Kelvin,
while the density reaches $n\sim 10^{20}\; {\rm cm}^{-3}$.
The contraction of the central part now becomes very slow,
and hydrodynamic shocks are generated at the surface where
supersonic gas infall is suddenly stopped. 
This is the moment of birth of a protostar.
The protostar has a mass of just 0.01 solar masses. 
It has a radius of $\sim 5\times 10^{11}$ cm at its formation. 
The small mass is expected from the Jeans mass at the
final adiabatic phase. The central particle number density of
the protostar is $\sim 10^{21} {\rm cm}^{-3}$ and the temperature is
well above 10,000 Kelvin.

A long standing question is whether or not a primordial gas cloud 
experiences vigorous fragmentation during its
evolution\cite{Yoshii, Silk83}.  Cosmological simulations performed so far 
showed consistently that a single small proto-stellar
core is formed first\cite{ABN, YOH08}. It appears that gas cloud
fragmentation does not occur. The so-called chemo-thermal instability
has been studied in detail. 
The results from a semi-analytic calculation\cite{OY03} and from direct
three-dimensional simulations\cite{Y06, YOH08} show that, at
all evolutionary phases, the locally estimated growth time for perturbations is
longer than, or only comparable to, the local dynamical time for
collapse. Hence, the cloud core does not fragment into multiple clumps
by chemo-thermal instability, but instead its collapse is accelerated.
Fragmentation of the cloud during later proto-stellar evolution
has been examined\cite{Machida08, Clark08}. Intriguingly, it was shown that 
a bar or a disk structure can become unstable to yield binary 
or multiple systems. 
This is an important issue to be explored further by three-dimensional
simulations. The role of angular momentum and its transfer,
and the radiative feedback effects from the central protostar(s)
need to be studied.

\begin{figure}
\psfig{file=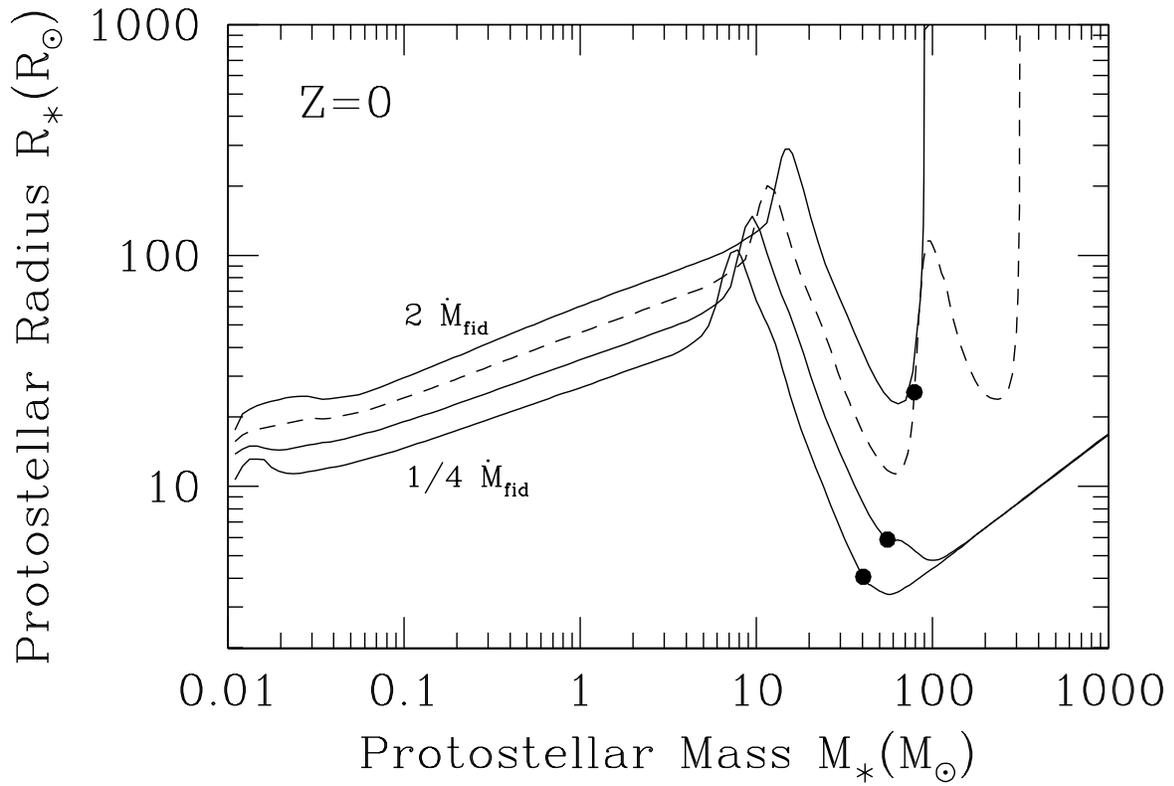,width=1\textwidth}
\caption{The evolution of the radius and mass of a primordial protostar.
The accretion rates assumed are 1/4, 1/2, 1, 2 $\dot{M}_{\rm fid}$
(from bottom to top)
with the fiducial rate of $\dot{M}_{\rm fid} = 4.4 \times 10^{-3} M_{\odot} {\rm yr}^{-1}$.
The solid points indicate the time when hydrogen burning begins.
From Ref. \cite{OP03}.
\label{fig:proto_evo}}
\end{figure}

On the assumption that there is only one stellar seed (protostar)
at the center of the parent gas cloud, the subsequent protostellar
evolution can be calculated using the standard model
of star formation\cite{Stahler86, OP01, OP03}.
For a very large accretion rate characteristic for a primordial gas cloud,
$\dot{M} > 10^{-3} M_{\odot}\;{\rm yr}^{-1}$, a protostar
can grow quickly to become a massive star.
Fig. \ref{fig:proto_evo} shows the evolution of proto-stellar
radius and mass for such large accretion rates.
The resulting mass when the star reaches the zero-age main sequence 
is as large as one hundred times that of the sun\cite{OP03, Y06}. 

Overall, the lack of vigorous fragmentation, the large gas mass
accretion rate, and the lack of significant source of opacity
(such as dust) 
provide favourable conditions for the formation
of massive, even very massive, stars in the early universe
\cite{Y06, McKee08, Ohkubo}.
A remaining important question is whether or not, and
how gas accretion is stopped. This question is directly related
to the final mass of the first stars.
A few mechanisms are suggested to act to stop gas accretion
and terminate the growth of a protostar\cite{McKee08}.
Following the growth of a primordial protostar to the end
of its evolution in a three-dimensional simulation will be 
the next frontier.

\section{Feedback from the first stars}
The birth and death of the first generation of stars have important
implications for the thermal state and chemical properties of the
intergalactic medium in the early universe.
At the end of the Dark Ages, the neutral, chemically pristine gas was
reionized by ultraviolet photons emitted from the first stars, but also
enriched with heavy elements when these stars ended their lives as
energetic supernovae.
The importance of supernova explosions, for instance, can be easily 
appreciated by noting that only light elements were produced during the 
nucleosynthesis phase in the early universe.  
Chemical elements heavier than lithium are thus thought to be
produced exclusively through stellar nucleosynthesis, and they must have
been expelled by supernovae to account for various observations of
high-redshift systems\cite{Walter, Songaila}.  

Feedback from the first stars may have played a crucial role
in the evolution of the intergalactic medium and (proto-)galaxy
formation. 
A good summary of the feedback processes is found in Ref. \cite{Ciardi}.
We here review two important effects, and highlight 
a few unsolved problems.

\subsection{Radiative feedback \label{sec:HII}}

The first feedback effect we discuss is radiation from the first stars.
First stars can cause both negative and positive 
-- in terms of star-formation efficiency -- feedback effects.
Far-UV radiation dissociates 
molecular hydrogen via Lyman-Werner
resonances\cite{Stecher67, Omukai99, Haiman00},
while UV photo-ionization heat up the surrounding gas.
Photo-ionization also increases the ionization fraction, which in turn promote
H$_{2}$ formation.
Yet another radiative feedback effect is conceivable; X-rays can promote
H$_2$ production by boosting the free electron fraction in distant
regions\cite{Oh01, Ricotti03}.  
It is not clear whether negative or positive feedback dominates
in the early universe.

\begin{figure}
\psfig{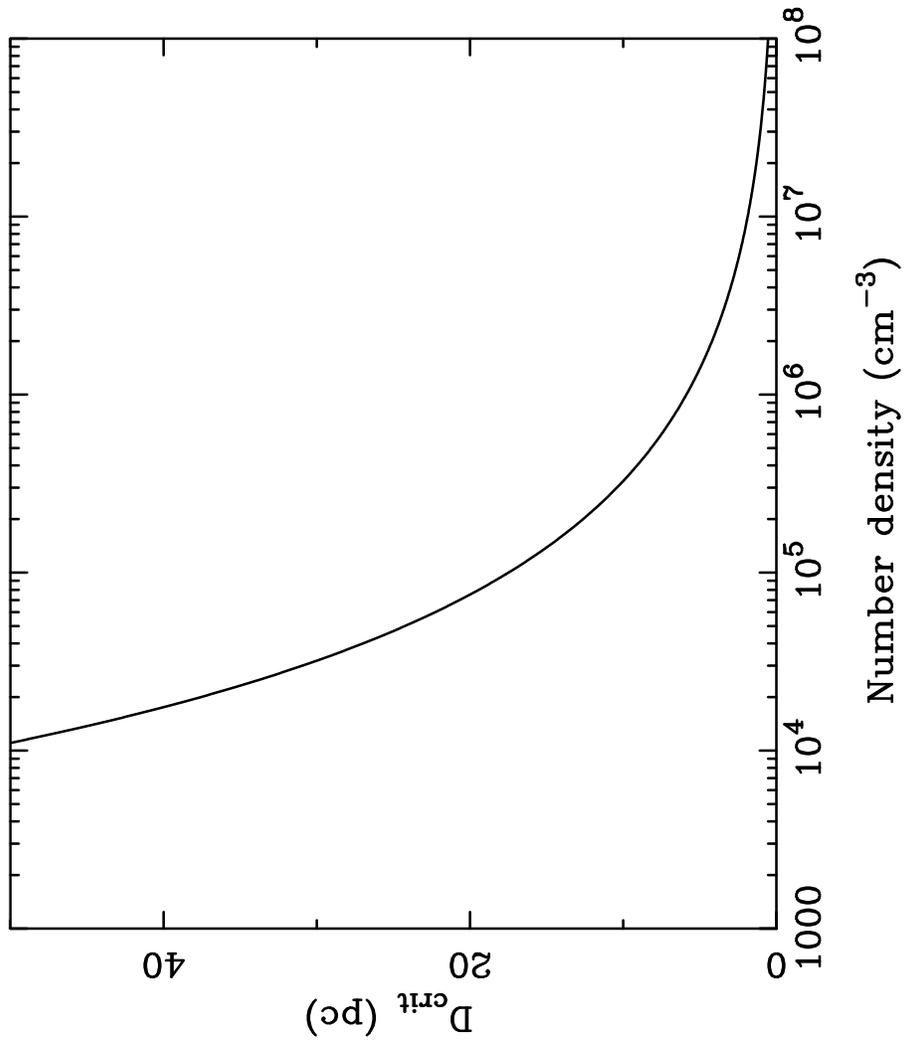}
\caption{The distance from a radiation source 
at which the dissociation and
free-fall timescales are equal, plotted as a function of
gas cloud density.
From Ref. \cite{Glover01}.}
\label{fig:ref}
\end{figure}

One-dimensional calculations (e.g. \cite{Glover01}) show consistently
strong negative effects of FUV radiation. 
Fig. \ref{fig:ref} shows the distance at which the 
H$_2$ dissociation time equals the free-fall time.
Hydrogen molecules in gas clouds within a few tens parsecs
are easily destroyed by a nearby massive star.
Three dimensional simulations also confirm the result
in the optically-thin limit\cite{MBA01}. 
However, gas self-shielding (opacity effects)
need to be taken into account for dense gas clouds.
H$_2$ dissociation becomes ineffective for large column densities of
$N_{\rm H_2} > 10^{14}$ cm$^{-2}$ for an approximately stationary gas\cite{Draine96}.
In fact, small halos are {\it not} optically-thin and thus the
gas at the center can be self-shielded against 
FUV radiations\cite{YoshidaFirst, Susa}.
Because of complexities associated with the dynamics, chemistry
and radiative transfer involved in early gas cloud formation, the strength of the 
radiative feedback still remains rather uncertain.
Recent simulations\cite{Ahn, Oshea} generally suggest that 
FUV radiation does not completely suppress star-formation
even for large intensities of $> 10^{-22}\, {\rm erg}\, {\rm sec}^{-1}\, {\rm Hz} \,{\rm cm}^{-2}$.
Contrary to the naive implication of the negative feedback
from FUV radiation, star-formation can possibly continue in early minihalos. 
In light of this, analytic and semi-analytic models need to be refined.
It is intriguing that the 5-year WMAP data
do not suggest a very large optical depth to Thomson scattering of CMB, 
perhaps constraining a large contribution to reionization from minihalos
\cite{Alvarez, HaimanB}.

If the formation of H$_{2}$ is strongly suppressed
by a FUV background, star formation proceeds in a quite 
different manner.
A primordial gas cloud cools and condenses nearly isothermally
by atomic hydrogen cooling. If the gas cloud has initially a small 
angular momentum, it can collapse 
to form an intermediate mass black hole may be formed\cite{BLBH, OmukaiH}.
Such first blackholes might power small quasars.
X-ray from early quasars is suggested as
a source of positive feedback effect by increasing the ionization fraction
in a primordial gas\cite{Ricotti03}.
However, the net effect is much weaker 
than one naively expects from simple analytic estimates
unless the negative feedback by FUV radiation is absent
\cite{MBA03}.

Ionizing radiation causes much stronger effects, at least locally.
The formation of early \HII regions were studied by a few groups using
radiation hydrodynamics simulations\cite{Kitayama04, Whalen04}.
Early \HII regions are different from present-day \HII regions
in two aspects.
Firstly, the first stars and their host gas cloud are hosted 
by a dark matter halo. Gravitational force exerted by dark matter
needs to be included in the dynamics of early \HII regions.
Secondly, the {\it initial} gas density profile around the first star
is typically steep\cite{ABN, Y06, YOH08}.
These two conditions make the evolution different
from that of present-day local \HII regions.

\begin{figure}
\psfig{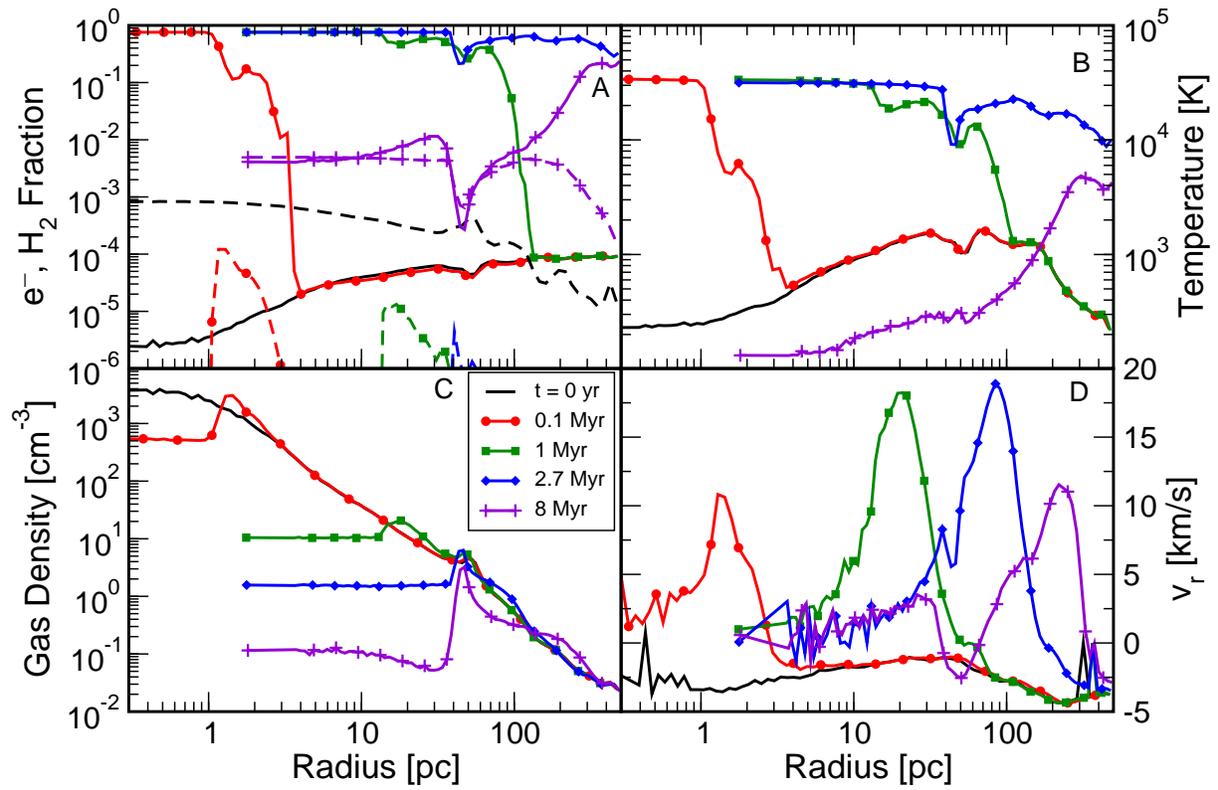}
  \caption{The structure and evolution 
of an \HII region around a massive Population III star
inside a minihalo. Radial profiles of ionization fraction, 
density, temperature, and
velocity are plotted for five output times. 
From Ref. \cite{AbelWise}}
  \label{fig:HII}
\end{figure}

 Fig.~\ref{fig:HII} shows the radial profiles of various quantities
in and around an early \HII region\cite{AbelWise}.
The star-forming region is located as a dense molecular gas cloud
within a small mass ($\sim 10^6 M_{\odot}$) dark matter halo.
A single massive Population III star with $M_{*}=100 M_{\odot}$ is embedded 
at the center.  
The formation of the \HII region is characterized by initial slow expansion 
of an ionization
front (I-front) near the center, followed by rapid propagation 
of the I-front throughout the outer gas envelope.  
The transition between the two phases
determines a critical condition for the complete ionization of the halo.  
For small mass halos, the 
transition takes place within a few $10^5$ years, and the I-front
expands over the halo's virial radius (Fig. \ref{fig:HII}).
The gas in the halo is effectively evacuated by a supersonic shock,
with the mean gas density decreasing to $\sim 1 {\rm cm}^{-3}$ 
in a few million years. 
It takes over tens to a hundred million years
for the evacuated gas to be re-incorporated in the halo\cite{YOKH07}. 
The most important implication from this result is that  
star-formation in the early universe would be intermittent.
Small mass halos can not sustain continuous star-formation.

Early gas clouds are expected to be strongly clustered\cite{Gao07, Reed}.  
Because even a single massive star affects over a kilo parsec volume, 
the mutual interactions between nearby star-forming gas clouds 
may be important.
Large-scale cosmological simulations with the radiative feedback effects 
such as those discussed here are clearly needed to fully explore the 
impact of early star formation.

\subsection{Mechanical feedback}
Massive stars end their lives as supernovae.
Such energetic explosions in the early universe are thought to be violently 
destructive; they expel the ambient gas out of the gravitational potential 
well of small-mass dark matter halos, causing an almost complete evacuation
\cite{BYH03, Wada03, Kitayama05, Greif07, Whalen08}.
Since massive stars process a substantial 
fraction of their mass into heavy
elements, SN explosions can cause prompt chemical enrichment, at least locally.
It may even provide an efficient mechanism to
pollute the surrounding intergalactic medium to an appreciable degree
\cite{Mori, YBH04}.  
 
Population III supernova explosions in the early
universe were also suggested as a trigger of star-formation\cite{CBA84},
but modern numerical simulations 
have shown that the expelled gas by supernovae
falls back to the dark halo potential well after about the system's  
free-fall time\cite{YBH04, johnson07}.
The density
and  density  profile  around  the  supernova sites  are  of  particular
importance  because the efficiency  of cooling  of supernova remnants 
is  critically determined by the density inside the blastwave.
If the halo gas is evacuated by radiative feedback prior to explosion, 
the supernova blastwave propagates over the halo's virial
radius, leading to complete evacuation of the gas even
with the input energy of $10^{51}$ erg. 
A large fraction of the remnant's thermal energy is lost in $10^5-10^7$ yr 
by line cooling, whereas, for even greater explosion energies, the remnant cools mainly 
via inverse Compton scattering. The situation is clearly different
from the local galactic supernova.
In the early universe, the inverse Compton process with cosmic background photons
acts as an efficient cooling process. 

\begin{figure}
\psfig{file=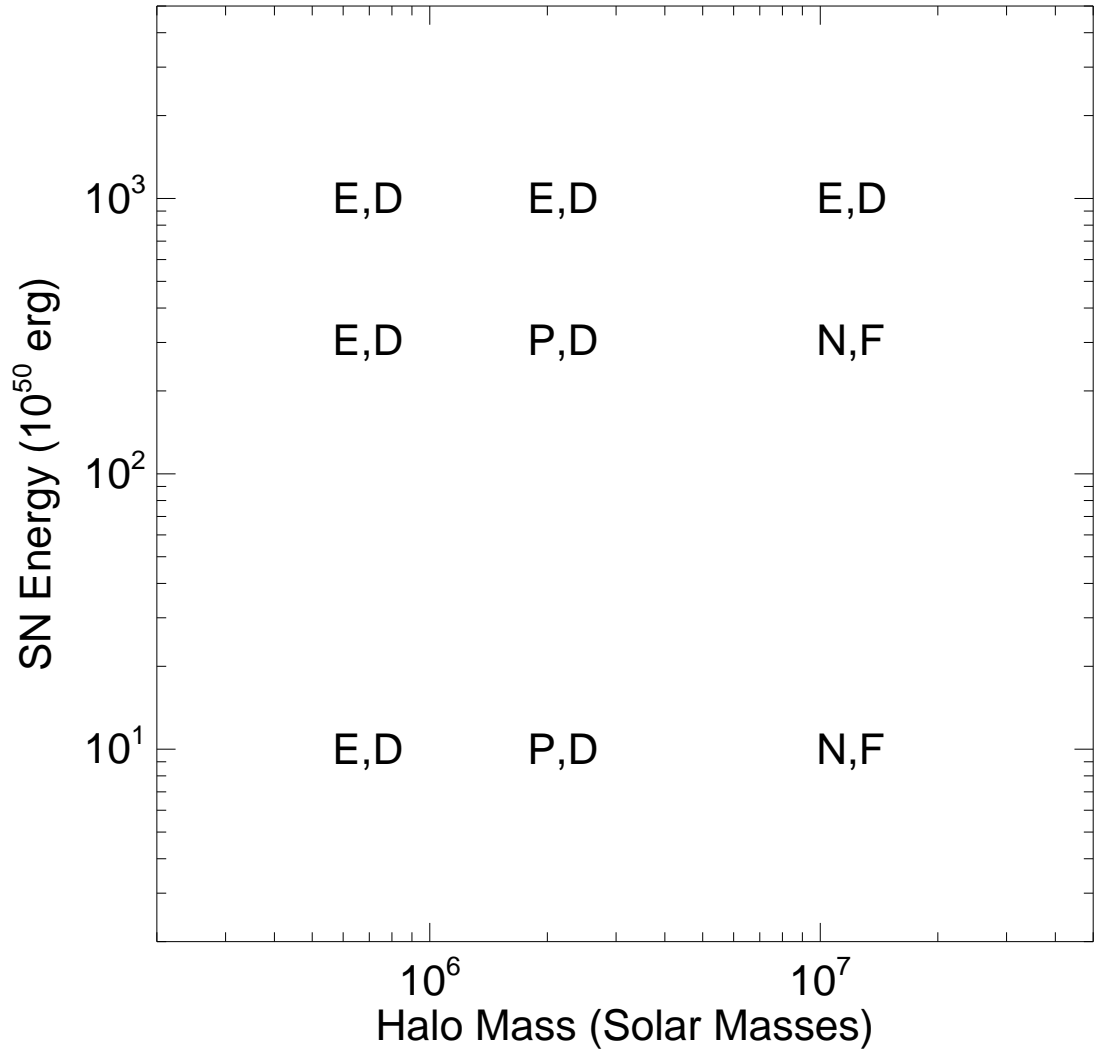,width=1\textwidth}
  \caption{Destruction efficiency of the first supernovae. 
The first letter refers to the ﬁnal state of the halo
prior to the explosion; E: photoevaporated; P: partly ionized, 
defined as the I-front not reaching the virial 
radius; N: neutral, or a failed \HII region. 
The second letter indicates outcome of the SN explosion; 
D: destroyed, or F: fallback. 
From Ref. \cite{Whalen08}.
}
\label{fig:sn}
\end{figure}

Fig.~\ref{fig:sn} summarizes the results from a series of calculations
of Ref. \cite{Whalen08}. It shows the destruction efficiency by 
a single SN explosion for a wide range of explosion energy and host
halo mass.
A simple criterion, $E_{\rm SN} > E_{\rm bi}$, where 
$E_{\rm bi}$ is the gravitational 
binding energy, is often used to determine the destruction efficiency.
However, whether or not the halo gas is effectively blown-away is determined not only
by the host halo mass (which gives an estimate of $E_{\rm bi}$), but also
by a complex interplay of hydrodynamics and radiative processes.
SNRs in dense environments are highly radiative and thus
a large fraction of the explosion energy can be quickly radiated away.
An immediate implication from the result is that, in order for the processed 
metals to be transported out of the halo and distributed to the IGM,
I-front propagation and pre-evacuation of the gas must precede the
supernova explosion. This roughly limits the mass of host halos from which 
metals can be ejected into the IGM to $<10^7 M_{\odot}$, i.e., the first generation
of stars can be a significant source of early metal-enrichment of the IGM
\cite{BYH03, YBH04, Greif07}.

Although metal-enrichment by the
first supernovae could greatly enhance the gas cooling efficiency, 
which might possibly change the mode of star-formation to that dominated by
low-mass stars\cite{schneider},
the onset of this `second-generation' stars may be delayed owing to
gas evacuation, particularly
in low-mass halos. This again supports the notion that 
early star-formation is likely self-regulating.
If the first stars are massive, only one period of star-formation is 
possible for a small halo and its descendants within a Hubble time.
The sharp decline in the destruction efficiency at 
$M_{\rm halo} >10^7 M_{\odot}$ (see Fig. \ref{fig:sn})
indicates that the global cosmic star formation activity increases only 
after a number of large mass ($> 10^{7-8} M_{\odot}$) halos are assembled.

\section{Toward the formation of the first galaxies}
The hierarchical nature of cosmic structure formation (see Section 2)
naturally predicts that stars or stellar size objects form first,
earlier than galaxies form. The first generation 
of stars set the scene for the 
subsequent galaxy formation. 
The characteristic minimum mass of a first galaxy (including dark matter)
is perhaps $\sim 10^{7}-10^{8} M_{\odot}$, in which the gas
heated up to $10^{4}-10^{5}$ Kelvin by the first star feedback
can be retained.

The first galaxies are assembled through a number of large 
and small mergers, and then turbulence is generated dynamically, which
likely changes star-formation process from a quiescent one (like in minihalos) 
to a highly complicated but organized one.
There have been a few attempts to directly simulate this
process in a cosmological context\cite{WiseAbel, Greif08}.
The results generally argue that star-formation in the
large mass system is still an inefficient process overall.
However, a significant difference is that the
inter-stellar medium is likely metal-enriched in the first galaxies.
Theoretical calculations\cite{BrommLoeb03, Omukai05}
show that cooling by heavy elements and by dust 
can bring the gas temperature at the onset of 
run-away collapse substantially lower than for
a primordial gas. The lower gas temperature
causes two effects; it lowers the Jeans mass $(\propto T^{3/2}/\rho^{1/2}$), and
also lowers the mass accretion rate $(\propto c_{\rm s}^{3}/G$),
thereby providing at least two necessary conditions for low-mass star-formation. 
Combined effects of strong turbulence and 
metal-enrichment might make the stellar initial
mass function be close to that in the present-day 
star-forming regions.

Understanding the formation of the first galaxies is
much challenging, because of the complexities described above.
Nevertheless it is definitely the subject where theoretical models can be really
tested against direct observations in the near future.
The first galaxies may be more appropriately called 
faint proto-galaxies, 
and will be detected by the next generation telescopes.
{\it JWST} will measure the luminosity function
of faint galaxies at $z > 7$, which reflects the strength of 
feedback effects from the first stars\cite{haiman08}.

\section{Prospects for future observations}
A number of observational programs are planed to detect
the first stars and galaxies, both directly and indirectly.
We close this review by discussing prospects for future
observations.

The first galaxies will be the main target of next generation
(near-)infrared telescopes, while indirect informations on 
the first stars will be obtained from the CMB polarization, 
the near-infrared background, high-redshift supernovae and gamma-ray bursts, 
and from the so-called Galactic archeology. 

The five-year data of the {\it Wilkinson Microwave Anisotropy 
Probe (WMAP)} yields the CMB optical depth to Thomson scattering,
$\tau\simeq 0.09\pm 0.03$\cite{Komatsu08}. 
This measurement provides an integral constraint on the total 
ionizing photon production at $z>6$ \cite{HuiH03}.
More accurate polarization measurements by {\it Planck}
and by a continued operation of
{\it WMAP} will further tighten the constraint on the reionization 
history of the universe, $x_{\rm e} (z)$\cite{Holder}.
In a longer term, 
future radio observations such as {\it Square Kilometer Array}
will map out the distribution of the intergalactic hydrogen
in the early universe.
The topology of reionization
and its evolution will be probed\cite{Shapiro06}. 

The first stars in the universe are predicted to be massive, 
as discussed in this article, and so they are likely progenitors of
energetic supernovae and associated GRBs at high redshifts\cite{WoosleyBloom}.
Infrared color can be utilized to identify supernovae at $z < 13$\cite{mesinger06}. 
A realistic 1-year {\it JWST} survey will discover 1-30 supernovae at $z > 5$\cite{haiman08}.
Gamma-ray bursts are the brightest explosions in the universe, and thus
are detectable out to redshifts $z>10$ in principle.
Recently, {\it Swift} satellite has detected a GRB originating at 
$z > 6$ \cite{Kawai, Greiner}, thus demonstrating the promise 
of GRBs as probes of the early universe.

Very metal-poor stars -- the stellar relics -- 
provide invaluable informations on the conditions under which these low-mass 
stars formed\cite{Christlieb, Frebel}. It is expected that the relics of early generation stars are orbiting
near the centers of galaxies at the present epoch\cite{White99}.
While conventionally halo stars are surveyed to find very metal-poor stars, 
the APOGEE project is aimed at observing $\sim 100,000$ stars
in the bulge of Milky Way\cite{apogee}. The nature of early metal-enrichment 
must be imprinted in the abundance patterns in the bulge stars.

Altogether, these observations will finally fill the gap in 
in our knowledge on the history of the universe,
and will end the ``Dark Ages''.

\bigskip
The work is supported in part by 
the Grants-in-Aid for Young Scientists (S) 20674003
by the Japan Society for the Promotion of Science.

\clearpage

\clearpage

\clearpage

\clearpage

\clearpage

\clearpage

\clearpage

\clearpage

\clearpage
\begin{figure}
\psfig{file=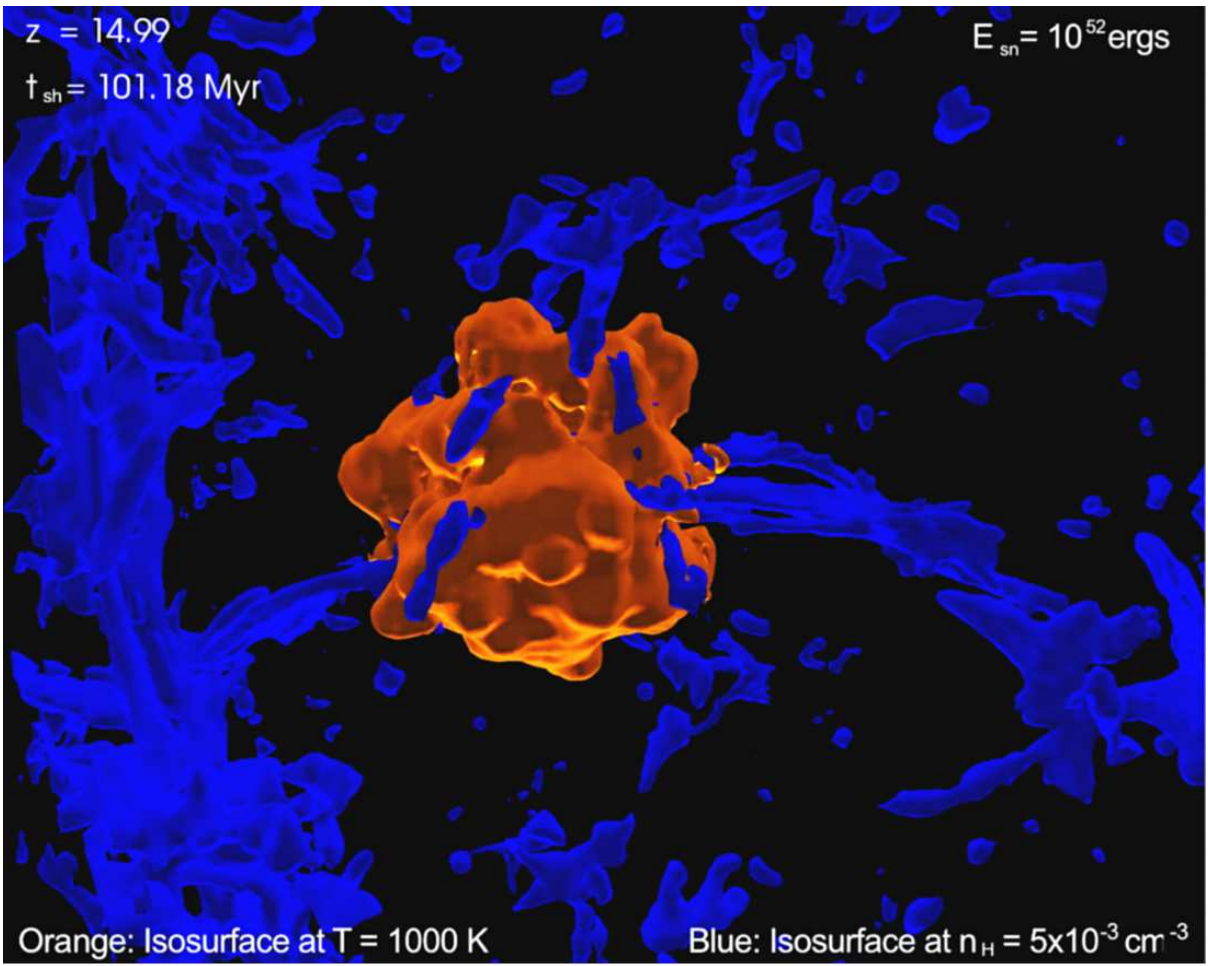,width=1\textwidth}
  \caption{The structure of an early 
supernova remnant. The shock-front reached a radius of 2 kpc
about 100 Myrs after the explosion. A large explosion energy of
$10^{52}$ ergs is assumed for this simulation.
From Ref. \cite{Greif08}.}
  \label{fig:sn3D}
\end{figure}

\end{document}